\documentclass[12pt]{article}

\usepackage{amssymb}
\usepackage{amsmath}

\begin{document} %


\title{
\bf General solution for the warp function in the RS scenario }

\author{
A.V. Kisselev\thanks{Electronic address:
alexandre.kisselev@ihep.ru} \\
{\small Institute for High Energy Physics, NRC ``Kurchatov
Institute''\!,} \\
{\small 142281 Protvino, Russian Federation}
}

\date{}

\maketitle

\thispagestyle{empty}


\begin{abstract}
The five-dimensional space-time, with non-factorizable geometry and
fifth dimension $y$ being an orbifold $S^1\!/Z_2$, is studied. In
such a scenario, originally suggested by Randall and Sundrum, there
exist two branes at fixed points of the orbifold, and the
four-dimensional metric is multiplied by a warp factor
$\exp[\sigma(y)]$. In the present paper, the general solution
$\sigma(y)$ of the Einstein-Hilbert's equations is presented which
is symmetric with respect to the interchange of two branes. It obeys
the orbifold symmetry $y \rightarrow - y$ and explicitly reproduces
jumps of its derivative on both branes. This general solution for
$\sigma(y)$ is determined by the Einstein-Hilbert's equations up to
a constant, that results in physically diverse schemes.
\end{abstract}





The 5-dimensional space-time with non-factorizable geometry and two
branes was suggested by Randall and Sundrum  (RS1 model)
\cite{Randall:99} as an alternative to the ADD model with flat extra
dimensions     \cite{Arkani-Hamed:98}-\cite{Arkani-Hamed:99}. Its
phenomenological implications were explored soon
\cite{Davoudiasl:00}. The model predicts an existence of heavy
Kaluza-Klein excitations (KK gravitons). These massive resonances
are intensively searched for by the LHC collaborations (see, for
instance, \cite{ATLAS:gravitons}, \cite{CMS:gravitons}).

The RS scenario is described by the following background warped
metric
\begin{equation}\label{RS_background_metric}
\quad ds^2 = e^{-2 \sigma (y)} \, \eta_{\mu \nu} \, dx^{\mu} \,
dx^{\nu} - dy^2 \;,
\end{equation}
where $\eta_{\mu\nu}$ is the Minkowski tensor with the signature
$(+,-,-,-)$, and $y$ is an extra coordinate. It is a model of
gravity in the AdS$_5$ space-time compactified to the orbifold
$S^1\!/Z_2$. There are two branes located at the fixed points of the
orbifold. The function $\sigma(y)$ in the warp factor $\exp[-2
\sigma(y)]$ was obtained to be \cite{Randall:99}
\begin{equation}\label{sigma_RS1}
\sigma_0 (y) = \kappa |y| \;,
\end{equation}
where $\kappa$ is a parameter with a dimension of mass.

This expression is consistent with the orbifold symmetry $y
\rightarrow - y$. However, it is not symmetric with respect to the
branes. The jump of the derivative $\sigma'(y)$ on the brane $y=\pi
r_c$ does not follow from expression \eqref{sigma_RS1}
\emph{directly}, but only after taking into account periodicity
condition.%
\footnote{Here and in what follows, the \emph{prime} denotes the
derivative with respect to variable $y$.}
Moreover, an arbitrary constant can be added to $\sigma_0
(y)$. Thus, a generalization of the RS solution \eqref{sigma_RS1} is
needed.

In the present paper we will derive such a general solution
$\sigma(y)$ of the Einstein-Hilbert's equations (see
\eqref{sigma_deriv_eq}, \eqref{sigma_2nd_deriv_eq} below) which has
the following properties: (i) it has the orbifold symmetry $y
\rightarrow - y$; (ii) jumps of $\sigma'(y)$ are explicitly
reproduced on both branes; (iii) it is symmetric with respect to the
interchange of the branes; (iv) it includes a constant term.

Previously, the solution for $\sigma(y)$ was studied in
ref.~\cite{Kisselev:13}. In the present paper we reconsider and
strengthen arguments used in deriving solution, as well as correct
expressions for $\sigma'(y)$ and 5-dimensional cosmological constant
$\Lambda$ presented in \cite{Kisselev:13}.


The classical action of the Randall-Sundrum scenario
\cite{Randall:99} is given by
\begin{align}\label{action}
S &= \int \!\! d^4x \!\! \int_{-\pi r_c}^{\pi r_c} \!\! dy \,
\sqrt{G} \, (2 \bar{M}_5^3 \mathcal{R}
- \Lambda) \nonumber \\
&+ \int \!\! d^4x \sqrt{|g^{(1)}|} \, (\mathcal{L}_1 - \Lambda_1) +
\int \!\! d^4x \sqrt{|g^{(2)}|} \, (\mathcal{L}_2 - \Lambda_2) \;,
\end{align}
where $G_{MN}(x,y)$ is the 5-dimensional metric, with $M,N =
0,1,2,3,4$, $\mu = 0,1,2,3$, and $y$ is the 5-th dimension
coordinate of the size $\pi r_c$. The quantities
\begin{equation}
g^{(1)}_{\mu\nu}(x) = G_{\mu\nu}(x, y=0) \;, \quad
g^{(2)}_{\mu\nu}(x) = G_{\mu\nu}(x, y=\pi r_c)
\end{equation}
are induced metrics on the branes, $\mathcal{L}_1$ and
$\mathcal{L}_2$ are brane Lagrangians, $G = \det(G_{MN})$, $g^{(i)}
= \det(g^{(i)}_{\mu\nu})$.

The periodicity condition, $y = y \pm 2\pi r_c$, is imposed and the
points $(x_{\mu}, y)$ and $(x_{\mu}, -y)$ are identified. So, one
gets the orbifold $S^1/Z_2$. We consider the case with two 3-branes
located at the fixed points $y = 0$ (Plank brane) and $y = \pi r_c$
(TeV brane). The SM fields are constrained to the TeV (physical)
brane, while the gravity propagates in all spatial dimensions.

From action \eqref{action} 5-dimensional Einstein-Hilbert's
equations follow
\begin{align}\label{H-E_equation}
\sqrt{|G|} & \left( \mathcal{R}_{MN} - \frac{1}{2} \, G_{MN}
\mathcal{R} \right) = - \frac{1}{4 \bar{M}_5^3} \Big[ \sqrt{|G|} \,
G_{MN}  \Lambda
\nonumber \\
&+  \sqrt{|g^{(1)}|} \, g^{(1)}_{\mu\nu} \, \delta_M^\mu \,
\delta_N^\nu \, \delta(y) \, \Lambda_1 +  \sqrt{|g^{(2)}|} \,
g^{(2)}_{\mu\nu} \, \delta_M^\mu \, \delta_N^\nu \, \delta(y - \pi
r_c) \, \Lambda_2 \Big] \;.
\end{align}
In what follows, the reduced scales will be used:
$\bar{M}_{\mathrm{Pl}} = M_{\mathrm{Pl}} /\sqrt{8\pi} \simeq
2.4\cdot 10^{18} \ \mathrm{GeV}$, and $\bar{M}_5 = M_5 /(2\pi)^{1/3}
\simeq 0.54 \, \bar{M}_5$.

In order to solve Einstein-Hilbert's equations, it is assumed that
the background metric respects 4-dimensional Poincare invariance
\eqref{RS_background_metric}. After orbifolding, the coordinate of
the extra compact dimension varies within the limits $0 \leqslant  y
\leqslant \pi r_c$. Then the 5-dimensional background metric tensor looks like%
\footnote{We ignore the backreaction of the brane term on the
space-time geometry.}
\begin{equation}\label{cov_metric_tensor}
G_{M\!N} = \left(
  \begin{array}{cc}
  g_{\mu\nu} & 0 \\
    0 & -1 \\
  \end{array}
\right) \;,
\end{equation}
where $g_{\mu\nu} = \exp(-2 \sigma) \, \eta_{\mu\nu}$. For the
background metric, the Einstein-Hilbert's equations are reduced to
the following set of two equations
\begin{align}
6 \sigma'^2 (y) &= - \frac{\Lambda}{4 \bar{M}_5^3} \;,
\label{sigma_deriv_eq} \\
3\sigma''(y) &= \frac{1}{4 \bar{M}_5^3} \, [\Lambda_1 \, \delta(y) +
\Lambda_2 \, \delta(\pi r_c - y)] \;.
\label{sigma_2nd_deriv_eq}
\end{align}
In between the branes (i.e. for $0 < y < \pi r_c$) we get from
\eqref{sigma_2nd_deriv_eq} that $\sigma''(y) = 0$, that results in
$\sigma'(y) = \kappa$, where $\kappa$ is a scale with a dimension of
mass.

Let us define dimensionless quantities $\lambda$, $\lambda_1$ and
$\lambda_2$ ($\lambda > 0, \lambda_{1,2} \neq 0$),
\begin{equation}\label{small_lambdas}
\Lambda = -24 \bar{M}_5^3 \kappa^2 \! \lambda  \;, \quad
\Lambda_{1,2} = 12\bar{M}_5^3 \kappa \lambda_{1,2} \;.
\end{equation}
Then we obtain
\begin{align}
\sigma'^2 (y) &= \kappa^2 \lambda \;, \label{sigma_mod_eq_1} \\
\sigma''(y) &= \kappa [\lambda_1 \, \delta(y) + \lambda_2 \ \delta(y
- \pi r_c)] \;.
\label{sigma_mod_eq_2}
\end{align}
The quantity $\kappa$ defines a magnitude of the 5-dimensional
scalar curvature.

The branes must be treated on an \emph{equal} footing. Thus, the
function $\sigma(y)$ should be symmetric with respect to the
simultaneous replacements $|y| \rightleftarrows |y - \pi r_c|$,
$\lambda_1 \rightleftarrows \lambda_2$. For the interval $0
\leqslant y \leqslant \pi r_c$, the
solution of eq. \eqref{sigma_mod_eq_2} looks like%
\footnote{We omitted a term linear in $y$, since it explicitly
violets the orbifold symmetries.}
\begin{equation}\label{sigma_lambdas}
\sigma (y) = \frac{\kappa}{4} [ ( \lambda_1 - \lambda_2) (|y| - |y -
\pi r_c | ) +  ( \lambda_1 + \lambda_2) (|y| + |y - \pi r_c | ) ] +
\mathrm{\ constant} \;,
\end{equation}
where
\begin{equation}\label{lambda_relation_1}
\lambda_1 - \lambda_2 = 2 \;.
\end{equation}
Note that eq.~\eqref{lambda_relation_1} guarantees that $\sigma'(y)
= \kappa$ for $0 < y < \pi r_c$.

There are two possibilities:
\begin{itemize}
  \item brane tensions have the same sign  \\
The function $\sigma(y)$ should be symmetric with respect to the
replacement $|y| \rightarrow |y - \pi r_c|$, since under such a
replacement the branes are interchanged (the fixed point $y=0$
becomes the fixed point $y=\pi r_c$, and vice versa). Then one has
to put $\lambda_1 - \lambda_2 = 0$ that contradicts
eq.~\eqref{lambda_relation_1}. Thus, this case cannot be realized.

  \item brane tensions have the opposite signs \\
The warp function $\sigma(y)$ must be symmetric under the
simultaneous substitutions $|y| \rightarrow |y- \pi r_c|$, $\kappa
\rightarrow - \kappa$. Thus, one has to take
\begin{equation}\label{lambda_relation_2}
\lambda_1 + \lambda_2 = 0 \;.
\end{equation}
\end{itemize}

It follows from \eqref{lambda_relation_1}, \eqref{lambda_relation_2}
that the brane tensions are
\begin{equation}\label{brane_lambdas_solution}
\lambda_1 = - \lambda_2 = 1 \;.
\end{equation}
As a result, we come to the unique solution:
\begin{equation}\label{sigma_solution}
\sigma (y) = \frac{\kappa}{2} ( |y| - |y - \pi r_c | ) +
\frac{\kappa \pi r_c}{2} - C \;.
\end{equation}
The constants in \eqref{sigma_solution} are chosen in such a way
that one has
\begin{equation}\label{sigma_internal}
\sigma (y) = \kappa y - C
\end{equation}
within the interval $0 < y < \pi r_c$. Taking into account the
periodicity condition and orbifold symmetry, we put
\begin{equation}\label{C_limits}
0 \leqslant C \leqslant \kappa \pi r_c \;.
\end{equation}

It follows from Einstein-Hilbert's
eq.~\eqref{sigma_mod_eq_2},%
\footnote{It means that eq.~\eqref{sigma_deriv_solution} must be
valid for all RS-like solution.}
as well as from \eqref{sigma_solution}, that
\begin{equation}\label{sigma_deriv_solution}
\sigma'(y) = \frac{\kappa}{2} \, [ \varepsilon(y) - \varepsilon(y -
\pi r_c) ] \;.
\end{equation}
Let us stress that the domain of definition of the function
$\varepsilon(x)$ in \eqref{sigma_deriv_solution} \emph{must be
constrained} to the interval $-\pi r_c < x < \pi r_c$. Outside this
region, one has to use the periodicity condition first in order to
define $\sigma'(y)$
correctly.%
\footnote{As one has to do for expression \eqref{sigma_RS1}.}
In particular, it means
\begin{equation}\label{epsilon}
\varepsilon(-y_0) = - \varepsilon(y_0) = -1 \;, \quad
\varepsilon(-y_0 - \pi r_c) = \varepsilon(- y_0 + \pi r_c) = 1 \;,
\end{equation}
for $0 < y_0 < \pi r_c$. Then we find from
\eqref{sigma_deriv_solution}, \eqref{epsilon} that $\sigma'(-y) =
-\sigma' (y)$, as it should be for the derivative of the symmetric
function $\sigma(y)$, and eq.~\eqref{sigma_mod_eq_1} says that
\begin{equation}\label{lambda_solution}
\lambda = 1 \;.
\end{equation}
In initial notations,
\begin{align}
\Lambda &= -24 \bar{M}_5^3\kappa^2 \;, \label{Lambda_fine_tuning}
\\
\Lambda_1 &= - \Lambda_2 = 12 \bar{M}_5^3 \kappa  \;.
\label{Lambdas_fine_tuning}
\end{align}
The RS1 fine tuning relations look slightly different
\cite{Randall:99},
\begin{align}
\Lambda &= -24 \bar{M}_5^3\kappa^2 \;,
\label{RS1_Lambda_fine_tuning}
\\
\Lambda_1 &= - \Lambda_2 = 24 \bar{M}_5^3 \kappa \;.
\label{RS1_Lambdas_fine_tuning}
\end{align}

It can be seen that our solution $\sigma(y)$ \eqref{sigma_solution}
obeys $Z_2$ symmetry if one remembers the periodicity in variable
$y$. Let us underline that expression \eqref{sigma_solution} is
symmetric with respect to the branes. Indeed, under the replacement
$y \rightarrow \pi r_c - y$, the positions of the branes are
interchanged (the point $y=0$ becomes the point $y=\pi r_c$, and
vice versa), while under the replacement $\kappa \rightarrow -
\kappa$, the tensions of the branes \eqref{Lambdas_fine_tuning} are
interchanged.

If we start from the fixed point $y = \pi r_c$ instead of the point
$y=0$, we come to an \emph{equivalent} solution
\begin{equation}\label{sigma_RS_pi}
\sigma_\pi(y) = -\kappa |y - \pi r_c| + \kappa \pi r_c \;.
\end{equation}
Note that \eqref{sigma_RS_pi} and \eqref{sigma_RS1} coincide at $0 <
y < \pi r_c$. Our final formula \eqref{sigma_solution} is in fact a
half-sum of these two solutions,
\begin{equation}\label{}
\sigma(y) = \frac{1}{2}[\sigma_0(y) + \sigma_\pi(y)] +
\mathrm{constant} \;.
\end{equation}

An explicit expression which makes the jumps of $\sigma'(y)$ on both
branes was presented in \cite{Dominici:03},
\begin{equation}\label{sigma_two_jumps}
\sigma(y) = \kappa \{y [2\,\theta(y) - 1] - 2 (y - \pi
r_c)\,\theta(y - \pi r_c) \}  + \mathrm{constant} \;.
\end{equation}
However, contrary to our formula \eqref{sigma_solution}, this
expression is neither symmetric in variable $y$ nor invariant with
respect to the interchange of the branes.

Let us stress that not only the brane warp factors, but hierarchy
relations and graviton mass spectra depend drastically on a
particular value of the constant $C$ in \eqref{sigma_solution}.
Correspondingly, the parameters of the model, $\bar{M}_5$ and
$\kappa$, can differ significantly for different $C$.

From now on, it will be assumed that $\pi\!\kappa r_c \gg 1$. The
hierarchy relation is given by the formula
\begin{equation}\label{hierarchy_relation}
\bar{M}_{\mathrm{Pl}}^2  = \frac{\bar{M}_5^3}{\kappa} e^{2C} \left(
1 - e^{-2\pi \kappa r_c} \right) \simeq \frac{\bar{M}_5^3}{\kappa}
\, e^{2C} \;.
\end{equation}
The interactions of the gravitons $h_{\mu\nu}^{(n)}$ with the SM
fields on the physical brane (brane 2) are given by the effective
Lagrangian
\begin{equation}\label{Lagrangian}
\mathcal{L}_{\mathrm{int}} = - \frac{1}{\bar{M}_{\mathrm{Pl}}} \,
h_{\mu\nu}^{(0)}(x) \, T_{\alpha\beta}(x) \, \eta^{\mu\alpha}
\eta^{\nu\beta} - \frac{1}{\Lambda_\pi} \sum_{n=1}^{\infty}
h_{\mu\nu}^{(n)}(x) \, T_{\alpha\beta}(x) \, \eta^{\mu\alpha}
\eta^{\nu\beta} \;,
\end{equation}
were $T^{\mu \nu}(x)$ is the energy-momentum tensor of the SM
fields, and the coupling constant of the massive modes is
\begin{equation}\label{Lambda_pi}
\Lambda_\pi \simeq \frac{\bar{M}_{\mathrm{Pl}}}{\sqrt{\exp(2\kappa
\pi r_c) - 1}}\simeq \bar{M}_{\mathrm{Pl}} \, e^{-\kappa \pi r_c}
\;.
\end{equation}

The graviton masses $m_n$ ($n = 1, 2, \ldots $) are defined from the
equation
\begin{equation}\label{masses_eq}
J_1 (a_{1n}) Y_1(a_{2n}) - Y_1 (a_{1n}) J_1(a_{2n}) = 0 \;,
\end{equation}
where
\begin{equation}\label{a_i}
a_1 = \frac{m_n}{\kappa} \, e^{-C} =
\frac{m_n}{\bar{M}_{\mathrm{Pl}}} \left( \frac{\bar{M}_5}{\kappa}
\right)^{3/2} \!, \quad a_2 = \frac{m_n}{\kappa} \, e^{\kappa \pi
r_c - C} = \frac{m_n}{\bar{M}_{\mathrm{Pl}}} \left(
\frac{\bar{M}_5}{\kappa} \right)^{3/2} \!\! e^{\kappa \pi r_c} \;.
\end{equation}
As a result, for all $m_n \ll  \bar{M}_{\mathrm{Pl}}
(\kappa/\bar{M}_5)^{3/2}$, we get
\begin{equation}\label{m_n}
m_n = x_n \bar{M}_{\mathrm{Pl}} \! \left( \frac{\kappa}{\bar{M}_5}
\right)^{3/2} \!\! e^{-\kappa \pi r_c} \;,
\end{equation}
where $x_n$ are zeros of the Bessel function $J_1(x)$.

By taking different values of $C$ in eq.~\eqref{sigma_solution}, we
come to quite diverse \emph{physical scenarios}. One of them ($C =
0$) is in fact the RS1 model \cite{Randall:99}. Another scheme ($C =
\kappa \pi r_c$) describes a geometry with a small curvature of
five-dimensional space-time \cite{Giudice:05}-\cite{Kisselev:06}
(RSSC model). It predicts a spectrum of the KK gravitons similar to
a spectrum of the ADD model
\cite{Arkani-Hamed:98}-\cite{Arkani-Hamed:99}. For the LHC
phenomenology of the RSSC model, see, for instance,
\cite{Kisselev:diphotons}, \cite{Kisselev:dimuons}. The scheme with
$C = \kappa \pi r_c/2$, and $\sigma (0) = - \sigma (\pi r_c) = -
\kappa \pi r_c/2$ also lead to an interesting phenomenology quite
different from that of the RS1 model. The details is a subject of a
separate publication.

As one can see, both the mass spectrum of the KK gravitons
\eqref{m_n} and theirs interaction with the SM fields
\eqref{Lambda_pi} is independent of $C$. However, it does not mean
that schemes  with different values of $C$ are physically
equivalent. The point is that the hierarchy relation
\eqref{hierarchy_relation} does depend on $C$. The RS1 hierarchy
relation looks like
\begin{equation}\label{hierarchy_relation_RS1}
\bar{M}_{\mathrm{Pl}}^2  = \frac{\bar{M}_5^3}{\kappa}  \;,
\end{equation}
while the RSSC relation \cite{Giudice:05}-\cite{Kisselev:06},
\cite{Rubakov:01} is
\begin{equation}\label{hierarchy_relation_RSSC}
\bar{M}_{\mathrm{Pl}}^2  = \frac{\bar{M}_5^3}{\kappa} \, e^{2\kappa
\pi r_c} \;.
\end{equation}

The different values of the constant $C$ leads to a quite different
spectra of the KK gravitons. For instance, in the RS1 model the
hierarchy relation \eqref{hierarchy_relation_RS1} needs $\kappa \sim
\bar{M}_5 \sim \bar{M}_{\mathrm{Pl}}$ with $m_n/x_n \sim 1$ TeV,
while in the RSSC model one can take $\kappa \sim 1$ GeV, $M_5 \sim
1$ TeV, that results in $m_n/x_n \sim 1$ GeV. Let us underline that
eq.~\eqref{hierarchy_relation_RS1} \emph{does not admit} the
parameters of the model to lie in the mentioned above region $\kappa
\sim 1$ GeV, $M_5 \sim 1$ TeV.

Note that a shift $\sigma \rightarrow \sigma - B$, where $B$ is a
constant, is equivalent to a change of four-dimensional
coordinates~\cite{Rubakov:01},
\begin{equation}\label{x_transformation}
x^\mu \rightarrow x'^{\mu} = e^{-B} x^\mu \;.
\end{equation}
However, a massive theory \emph{is not} invariant with respect to
scale transformations. Indeed, consider the effective gravity action
on the TeV brane (with radion term omitted). It looks like (see, for
instance, \cite{Boos:02})
\begin{equation}\label{gravity_TeV_action}
S_{\mathrm{eff}} = \frac{1}{4} \sum_{n=0}^{\infty} \int \! d^4x \!
\left[ \partial_\mu h^{(n)}_{\varrho\sigma}(x) \partial_\nu
h^{(n)}_{\delta \lambda} (x) \, \eta^{\mu\nu} - m_n^2
h^{(n)}_{\varrho\sigma} (x) h^{(n)}_{\delta \lambda}(x) \right] \!
\eta^{\varrho\delta} \eta^{\sigma\lambda} \;,
\end{equation}
The invariance of this action under transformation
\eqref{x_transformation} needs rescaling of the graviton fields and
their mass: $h^{(n)}_{\mu\nu} \rightarrow h'^{(n)}_{\mu\nu} = e^{B}
h^{(n)}_{\mu\nu}$, $m_n \rightarrow m'_n = e^{B} m_n$. The scale
transformation \eqref{x_transformation} is a particular case of
conformal transformations, and the theory of \emph{massive} KK
gravitons is not conformal invariant. The zero mass sector (standard
gravity) remains the same under the scale transformation.

As an illustration, note that the transition from the RS1 scenario
to the RSSC scenario means that $\sigma \rightarrow \sigma -
\pi\kappa r_c$. Correspondingly, equation for the graviton masses in
the RS1 model,
\begin{equation}\label{masses_RS1}
m_n \simeq x_n \kappa \, e^{-\kappa\pi r_c} \;,
\end{equation}
transforms into the following equation (see also
\cite{Giudice:05}-\cite{Kisselev:06}, \cite{Rubakov:01})
\begin{equation}\label{masses_RSSC}
m_n \simeq x_n \kappa\;.
\end{equation}

In the limit $\kappa \rightarrow 0$, the hierarchy relation for the
flat metric is reproduced from \eqref{hierarchy_relation},
\begin{equation}\label{flat_hierarchy_relation}
\bar{M}_{\mathrm{Pl}}^2  = \bar{M}_5^3 V_1 \;,
\end{equation}
where $V_1 = 2\pi r_c$ is the ED volume.%
\footnote{Note that $C \rightarrow 0$ in this limit, since $0
\leqslant C \leqslant \kappa \pi r_c$.}
In addition, $\Lambda_\pi \rightarrow \bar{M}_{\mathrm{Pl}}$, and
$m_n \rightarrow n/r_c$, as one can derive from \eqref{masses_eq}.



To summarize, we have studied the space-time with non-factorizable
geometry in four spatial dimensions with two branes (RS scenario).
It has the warp factor $\exp[\sigma(y)]$ in front of
four-dimensional metric. The generalization of the original RS
solution of the Einstein-Hilbert equations for the function
$\sigma(y)$ is obtained \eqref{sigma_solution} which: (i) obeys the
orbifold symmetry $y \rightarrow - y$; (ii) makes the jumps of
$\sigma'(y)$ on both branes; (iii) has the explicit symmetry with
respect to the branes; (iv) includes the constant $C$ ($0 \leqslant
C \leqslant \kappa \pi r_c$). This constant can be used for model
building within the framework of the general RS scenario.



The author is indebted to G.~Altarelli, I.~Antoniadis, M.L. Mangano
and V.A.~Petrov for fruitful discussions. He is also indebted to the
Theory Division of CERN for the support and hospitality.




\end{document}